\documentclass[12pt,preprint]{aastex}

\newcommand{\kms}{km~s$^{-1}$}

\shorttitle{\ion{H}{1} clouds in the M31 halo}
\shortauthors{Thilker et al.}

\begin{document}


\title{On the continuing formation of the Andromeda Galaxy:\\
Detection of \ion{H}{1} clouds in the M31 halo}


\author{David A. Thilker,\altaffilmark{1} Robert Braun,\altaffilmark{2} Ren\'e A. M. Walterbos,\altaffilmark{3} Edvige Corbelli,\altaffilmark{4}\\
Felix J. Lockman,\altaffilmark{5} Edward Murphy,\altaffilmark{6} Ronald Maddalena\altaffilmark{5}}


\altaffiltext{1}{Center for Astrophysical Sciences, Johns Hopkins University,
3400 North Charles Street, Baltimore, MD 21218 USA}

\altaffiltext{2}{Netherlands Foundation for Research in Astronomy,
P.O. Box 2, 7990 AA Dwingeloo, The Netherlands}

\altaffiltext{3}{Department of Astronomy, New Mexico State University,
MSC 4500, Box 30001, Las Cruces, NM 88003 USA}

\altaffiltext{4}{INAF-Osservatorio Astrofisico di Arcetri,
Largo E. Fermi 5, I-50125, Firenze, Italy}

\altaffiltext{5}{National Radio Astronomy Observatory$^0$,
P.O. Box 2, Green Bank, WV 24944-0002 USA}

\altaffiltext{6}{Department of Astronomy, University of Virginia,
P.O. Box 3818, Charlottesville, VA 22903-0818 USA}

\altaffiltext{7}{The
National Radio Astronomy Observatory is a facility of the National
Science Foundation operated under cooperative agreement by Associated
Universities, Inc.}


\begin{abstract}

Green Bank Telescope (GBT) $\lambda$21cm observations have revealed a
faint, yet extensive \ion{H}{1} cloud population surrounding the
Andromeda Galaxy (M31).  The newfound objects are likely analogs to
the high-velocity \ion{H}{1} clouds (HVCs) seen around the Milky Way.
At least 20 discrete features are detected within 50~kpc of the M31
disk, with radial velocities that are comparable to those of outer
disk rotation. In addition, a filamentary ``halo'' component of at
least 30~kpc extent is concentrated at the M31 systemic velocity. Some
of the discrete features are organized into elongated systems with
velocity continuity, suggestive of tidal streams. The discrete
population can be characterized by a steep power-law distribution of
number versus \ion{H}{1} mass in the range between 10$^5$ and
10$^7$~M$_\odot$. The velocity line-width of discrete clouds is
correlated with the cloud \ion{H}{1} mass: such that if the clouds are
gravitationally bound this implies a dark- to \ion{H}{1} mass ratio of
$\sim$ 100:1. Possible origins for the discrete and ``halo'' M31
features include: a Local Group ``cooling flow'', tidal debris from
recent mergers or interactions, and the gaseous counterparts of low
mass dark-matter halos.

\end{abstract}







\section{Introduction}

Much of the sky is covered with neutral hydrogen clouds whose
velocities, though within several hundreds of km s$^{-1}$ of zero, are
nonetheless anomalous and cannot be explained by normal rotation of
the Galaxy \citep{mulleretal63, wvw97}.  It is attractive to consider
that these high-velocity HI clouds (HVCs) might be material left over
from the formation of the Milky Way, now in the process of being
accreted \citep{oort70}.  If this is the case, other galaxies may be
expected to have similar gaseous components.  A key advantage of
studying such extragalactic clouds is that their distance can be
comparatively well-constrained. In practice, it has proven difficult
to achieve the combination of high spatial resolution, mass
sensitivity, and large field-of-view necessary to detect HVC analogs
in external systems.  Many nearby galaxies have a few peculiar
\ion{H}{1} features with \ion{H}{1} mass as low as $\sim$
10$^7$~M$_\odot$ (eg. NGC~628, \citet{kamphuisbriggs92}), but only
recently have instruments had sufficient sensitivity to begin
revealing complete systems of peculiar velocity \ion{H}{1} (eg.  the
slowly rotating ``halo'' component of NGC~2403,
\citet{fraternalietal02}).  In this Letter, we report on deep,
wide-field observations conducted with the Green Bank Telescope (GBT)
which have resulted in the detection of a complex system of \ion{H}{1}
gas extending to at least 50~kpc radius around M31.

\section{Observations and Data Analysis}

The GBT observations were obtained over six nights during September
2002.  The telescope was scanned over a $7\arcdeg \times 7\arcdeg$
region around M31 while \ion{H}{1} spectra were measured over a
heliocentric velocity range between $-827$ \kms\ and +226 \kms\ at a
velocity resolution of 1.25 \kms.  At M31's distance of 770 kpc
\citep{freedmanmadore90}, the GBT beam (9.1$\arcmin$ FWHM) projects to
2.0 kpc and our survey covers a $94\times94$~kpc$^2$ region.  At a
resolution of 3 kpc and 18 \kms\, the acheived rms flux sensitivity
was 7.2 mJy beam$^{-1}$ corresponding to an \ion{H}{1} column density
of $2.5\times10^{17}$~cm$^{-2}$ in a single channel averaged over the
beam.

Our GBT data show many \ion{H}{1} clouds surrounding M31 at $-515 \leq
V_{hel} \leq -172$ \kms, which can be compared to the systemic
velocity of $-300$ \kms.  Emission features were identified as
discrete sources if they were distinct from the M31 \ion{H}{1} disk
(which extends over the velocity interval $-620$ to $-20$ \kms), were
unconfused with the Galaxy, and were characterized by a peak flux
density $ >5\sigma$.  Additional GBT spectra were obtained for the
faintest detections to verify the accuracy of our procedures.  We
limited the investigation to velocities more negative than $-160$
\kms\ in order to exclude contamination from intermediate velocity
Galactic emission.

\section{\ion{H}{1} Clouds Around M31}

Our $\lambda21$cm observations have revealed a population of about
twenty discrete clouds together with an extended, filamentary ``halo''
component.  The data show that these features cannot be instrumental
effects related to M31's bright \ion{H}{1} disk, for there is no
relationship between the diffuse structures and the brightness of
nearby disk gas.  GBT images showing the distribution of \ion{H}{1} at
different velocities are presented in Fig. 1. These were taken from a
heavily-smoothed variant of the data cube.  Contours are drawn at 
2, 4, and 6$\sigma$.  Discrete compact \ion{H}{1} clouds appear at
velocities between $-172$ and $-515$ \kms.  Extended regions of
filamentary, low N$_{\rm HI}$ gas are found preferentially within
$\sim80$ \kms\ of M31's systemic velocity ($\rm{v}_{hel} = -300$
\kms).

We believe that this gas is associated with M31, for the following
reasons: (1) there is evidence for interaction between several of the
clouds and M31 or its companions, (2) cloud velocities match the
velocity extent of M31's disk and generally correlate with the pattern
of Andromeda's outer disk rotation, (3) the extended \ion{H}{1}
component is concentrated near M31's systemic velocity, (4) confusion
with the foreground Magellanic Stream at this location is unlikely,
and (5) dynamical constraints on the mass of M31, assuming a bound
cloud population, are in good agreement with independently determined
values.  We first examine this evidence linking the \ion{H}{1}
detections to M31, then consider possible origins for the cloud
population.

Figure 2 depicts the high velocity gas in the survey field which could
be cleanly separated from the disk emission of Andromeda. Contours of
integrated \ion{H}{1} column density at 0.5, 1, 2, 5, 10 and
$20\times10^{18}$~cm$^{-2}$ (evaluated using the smoothed cube shown
in Fig. 1) are overlaid on an optical image of M31.  Star
symbols mark discrete clouds meeting our 5$\sigma$
significance threshold which are adequately resolved from neighbors
in position-velocity space.  One discrete cloud near the north edge of
the field peaks below the minimum contour after datacube smoothing,
but has been independently confirmed in follow-up GBT observations.
In addition to discrete clouds, there are extended filamentary
complexes of low column density \ion{H}{1}, most notably at small
galactocentric radii southeast and north of Andromeda's stellar disk.
Only structures which are cleanly separated from the M31
disk \ion{H}{1} emission are included in Fig. 2. Additional filaments
can be discerned in Fig. 1.

The collection of discrete \ion{H}{1} clouds extending more than
$1.5\arcdeg$ (20~kpc) to the south of M31's disk near
($\alpha_{J2000}$, $\delta_{J2000}$) = (00:42, 39:15) is suggestive of
a tidal stream, based on the elongated distribution and continuously
varying radial velocity of the features. These \ion{H}{1} clouds are
partially coextensive with the metal-rich stellar ``Andromeda Stream''
\citep{ibataetal01,irwinetal02,fergusonetal02,mcconnachieetal03}
although they are significantly offset to the southwest.  The
conclusion that these particular clouds are associated with M31 seems
inescapable. A second case of partial correspondence of gaseous and
stellar components is the discrete object which is displaced by only
25$\arcmin$ to the southwest of NGC 205 and is overlapping in velocity
with that galaxy. This feature also merges in position and velocity
with the M31 disk.

The only object in our sample that was known previously is Davies'
Cloud \citep{davies75}, which is at least 10 times brighter than the
other features. Davies considered that it might be part
of the Magellanic Stream rather than M31, but in our complimentary
wide-field study of a $60\arcdeg\times 30\arcdeg$ region centered on
M31 (\citet{braunetal03}; see also Braun \& Thilker 2003, in prep.)
we find that the Magellanic Stream comes no closer than $7\arcdeg$ to
the southwest, where its velocity differs by about 100 \kms\ from the
diffuse components near M31.  It seems likely that Davies' Cloud and
the other detections from our survey are not confused by emission from
the Magellanic Stream.  A high resolution imaging study of Davies'
Cloud \citep{deheijetal02a} has provided additional morphological and
kinematic evidence for tidal interaction with M31. We suggest that it
is simply the most massive of a population of faint \ion{H}{1} clouds
around the Andromeda Galaxy.

Additional evidence for association of the newly-detected clouds 
with M31 is that the 
discrete cloud velocities appear partially correlated with the pattern
of outer disk rotation in M31, with the most negative velocities occurring in
the south-west and the most positive in the north-east. 
The high negative velocities of some clouds  (e.g., $-$515 \kms) 
are also more extreme than any seen previously in Galactic HVC surveys
($\rm{v} > -466$ \kms) \citep{putmanetal02,deheijetal02b}. 
 A straightforward interpretation is that the clouds are collectively
under the gravitational influence of the Andromeda Galaxy and have 
interacted dissipatively with M31 in the past.

If the detected clouds represent a circum-galactic population
gravitationally bound to Andromeda, their velocity offsets and
projected separations should be collectively related to M31's total
mass.  Using the virial theorem and representing the discrete GBT
clouds as bound test particles orbiting a central mass \citep[Eqn
10-22]{binneytremaine87}, we find that M$_{M31} =
5.2\times10^{11}$~M$_{\odot}$.  This estimate reflects only the mass
contained within the radii probed by our clouds.  If we instead assume
an extended mass distribution (Eqn 10-23) similar to the distribution
of clouds, the entire virial mass of Andromeda is then M$_{M31,vir}$ =
$7.9\times10^{11}$~M$_{\odot}$.  These two mass determinations are in
good agreement with independent total mass estimates of
$4\times10^{11}$~M$_{\odot}$ within a radius of 30 kpc
\citep{brinksburton84} and $1.6\times10^{12}$~M$_{\odot}$ within
Andromeda's virial radius of 300 kpc \citep{klypinetal02}.

From all these considerations, we consider it very likely that the
newfound \ion{H}{1} clouds reside in the circum-galactic environment
of M31.

\section{Discussion}

At a distance of 770 kpc, most clouds in our sample have \ion{H}{1}
masses in the range $0.15-1.3\times10^6$~M$_{\odot}$.  Such clouds
would not have been detected as discrete objects in previous surveys
of more distant galaxies, which generally have probed only the mass
range above $\sim10^7$~M$_{\odot}$ \citep{braunburton01}. An
indication for the more general existence of low mass circum-galactic
populations comes from the recent detection of a 15\% excess of
\ion{H}{1} mass found on scales of several hundred kpc relative to
that seen at tens of kpc in the environment of apparently isolated
galaxies \citep{braunetal03}.  Placing accurate limits on the total HI
mass of the M31 halo cloud population is difficult because we lack a
well-measured radial scale length for the population, must cope with
potential incompleteness due to Galactic confusion, and have no
confirmed minimum cloud mass.  To the degree that the observed cloud
mass function is consistent with a sensitivity-bounded power law given
by $dN(\rm{M}_{HI})~\alpha~\rm{M}_{HI}^{-2}~d\rm{M}_{HI}$, our
estimate of HI mass for the halo cloud population within the GBT field
is $\sim 3-4\times10^7$~M$_{\odot}$.  The upper limit of this range is
calculated assuming the true population is bounded by a minimum HI
cloud mass of $10^5$~M$_{\odot}$ and that there are fewer than $\sim
100$ of such objects, the vast majority of which are not yet detected.
We also included an extra 25\% to account for clouds confused with
Galactic emission, but even so, the \ion{H}{1} mass of Andromeda's
discrete halo cloud population in our data amounts to only $1\%$ of
the mass of M31's HI disk.  Below, we explore the idea that the halo
clouds may trace a more substantial amount of ionized gas and possibly
also dark matter.

An obvious source of high velocity \ion{H}{1} is tidal stripping from
recent or ongoing mergers.  We detect a gaseous feature which is
partially coextensive with the stellar Andromeda Stream
\citep{mcconnachieetal03}, and of comparable spatial extent, as well
as a component of possibly stripped gas adjacent to NGC~205.  However,
most of the discrete clouds detected in our GBT survey are rather
isolated, lacking any apparent relation to known M31 companion
galaxies, and do not have obvious indications if internal tidal
distortions.  Nevertheless, the faintest dwarf galaxies and stellar
streams are notoriously hard to detect (Armandroff et al. 1999). One
might argue that some dwarfs have yet to be discovered, while others
have been entirely consumed by M31.  Indeed, the comparatively young
(6-8 Gyr) halo stars found by Brown et al. (2003) in the ACS Andromeda
deep field do suggest that M31 has seen a major merger (with a massive
star forming galaxy) or several minor mergers.

Cooling of a tenuous inter-galactic medium \citep{oort70} is a second
viable source of clouds.  In this view, halo clouds would condense
from, and remain confined by, coronal gas which is located around M31
or perhaps pervades the Local Group.  Sembach et al. (2002) presented
evidence for an extended and highly ionized Galactic corona or Local
Group medium traced by high-velocity \ion{O}{6} absorption.  Likewise,
Lockman et al. (2002) conducted a sensitive HI survey of 860
sightlines at $\delta > -43\arcdeg$, which suggested that a low-column
density ``mist'' of high-velocity neutral gas surrounds the Milky Way
($\rm{N}_{HI} \ga 8\times10^{17}$~cm$^{-2}$ along $40\%$ of
sightlines).  Together, these studies show that the Galactic HVC
phenomenon extends to much lower column densities than traditionally
appreciated and that the classical Galactic HVCs are only part of a
more ubiquitous multi-phase medium. The filamentary ``\ion{H}{1}
halo'' we detect concentrated on the M31 systemic velocity may be a
manifestation of an M31 ``cooling flow''. In such a scenario, the
detected \ion{H}{1} may represent only the tip of the iceberg, in
terms of baryonic mass, since the gas may well be only of order 1\%
neutral at the relevant low volume densities \citep{sembachetal99}.
The primary source of such coronal gas may actually be the action of a
large-scale ``galactic fountain'' \citep{shapiro, bregman80,
savageetal03} in the recent or distant past.

A third component of high velocity \ion{H}{1} near M31 might be the
gas associated with a putative population of low mass dark-matter
halos. Current simulations of the Local Group in a $\Lambda$CDM
cosmology predict a large population of low mass DM halos at the
present epoch \citep{klypinetal99,mooreetal99} which dramatically
outnumber known dwarf galaxies. If the M31 halo clouds are tracers of
substantial dark-matter concentrations, this should be reflected in
their internal line-widths. We plot the observed \ion{H}{1} mass versus
FWHM line-width for discrete clouds in Figure 3.  Line-width
measurements were determined using Gaussian fits to cloud spectra.
The line-width distribution has an approximate lower bound roughly
consistent with thermal broadening for gas at 10$^4$ K (FWHM = 24
\kms). A systematic increase in FWHM line-width is observed with
increasing \ion{H}{1} mass, reaching some 70 \kms\ for the clouds near
10$^6$~M$_\odot$.  Although Davies' Cloud is significantly offset from
the remainder of the distribution in Fig. 3, high-resolution imaging
has revealed much higher internal line-widths in that object as well
\citep{deheijetal02a}. To demonstrate the expected distribution of
line-width with mass, in Fig.~3 we show a curve corresponding to
$V^{2} = 100 G M_{HI} R^{-1}$, where the characteristic discrete cloud
radius, R, has been held constant at 500 pc, based on our
interferometric WSRT detections of some of the cloud cores (Braun et
al. 2003, in prep.).  This curve, corresponding to a dark to
\ion{H}{1} mass ratio of about 100:1, is not intended to fit to the
data, but merely to provide a basis for comparison. The hypothesis of
a kinematically dominant dark-matter component appears to be
consistent with the observed line-widths of the discrete M31 halo
clouds.

Another aspect of the dark-matter mini-halo scenario that can be
checked is the expected number of such objects in the appropriate mass
range in the vicinity of M31. Sternberg et al. (2002) predict $\sim 25$
dark matter mini-halos associated with gravitationally confined HI
within a radius of 40 kpc around M31, based on the simulations of
Klypin et al. (1999) and Moore et al. (1999). This is fully consistent
with the 20 discrete M31 halo clouds we have detected.  The Sternberg et
al. calculations also suggest that circum-galactic objects of such low
mass and peak \ion{H}{1} column density should be only $\sim10$ \%
neutral, implying $M_{HI+HII} \sim 10 M_{HI}$ and $M_{DM} \sim 10
M_{HI+HII}$. Associated ionized gas could perhaps be
detected via deep Fabry-Perot imaging
\citep{blandhawthornmaloney99,tufteetal02} or absorption line spectroscopy of
background quasars \citep{murphyetal00,trippetal03,sembachetal03}.

In summary, our GBT observations of the Andromeda Galaxy have revealed
the first extensive extragalactic 
counterpart to the Galactic HVC population. Both discrete and
diffuse components are detected. We find some supporting evidence for
at least three different possible origins of the high velocity gas,
namely: tidal disruption, halo condensation and association with
low-mass DM halos.  

\clearpage


\begin{figure*}
\includegraphics[scale=0.6]{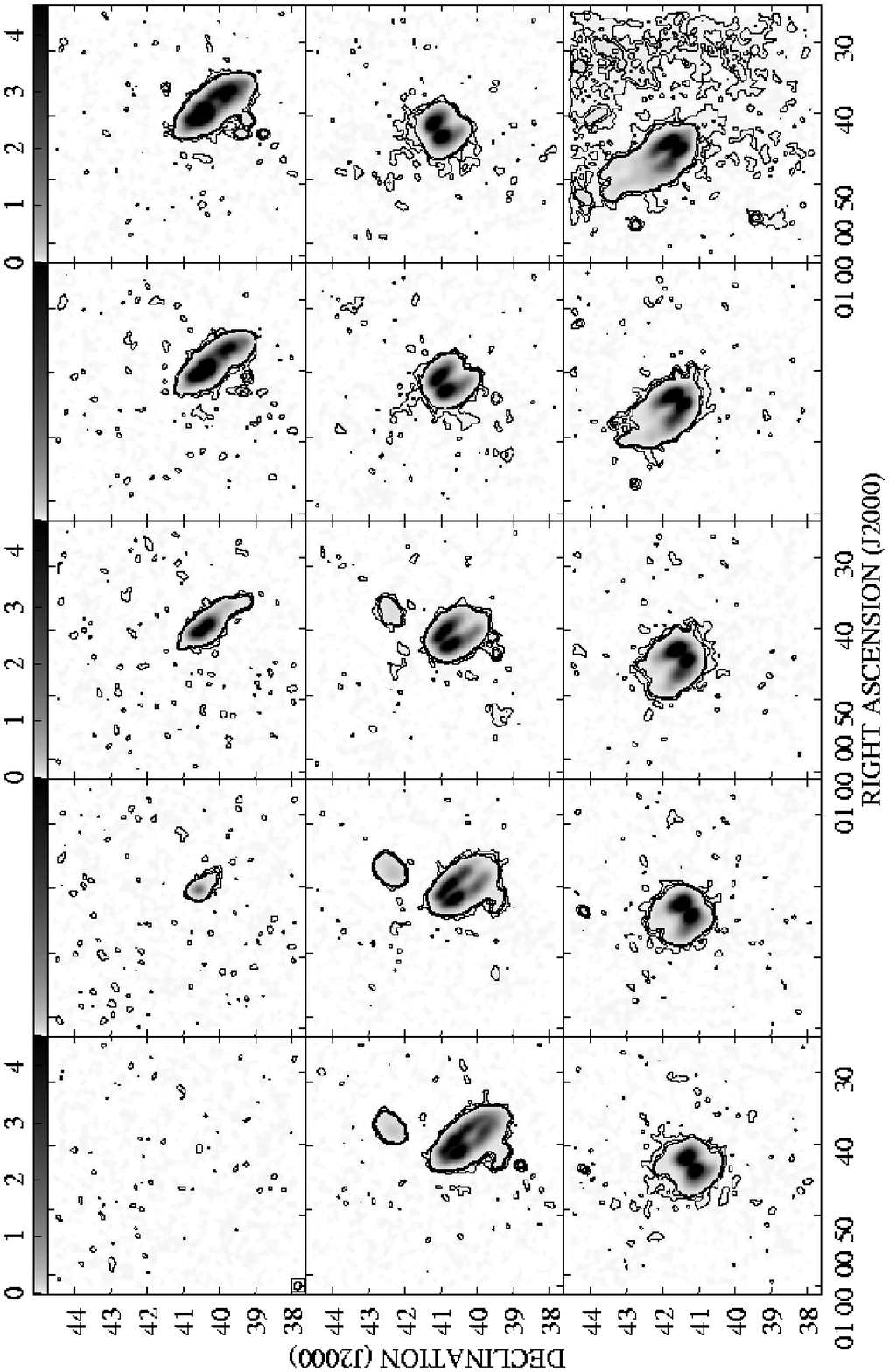}
\caption{ The overall distribution of high-velocity \ion{H}{1} gas in
the GBT survey volume, displayed with a square-root 
transfer function ($0-4.5$ Jy beam$^{-1}$).  Contours are
drawn at 2, 4, and 6$\sigma$.  Our data have been smoothed to (3
kpc, 72 \kms) resolution for this figure.  Note the population of
discrete high-velocity clouds, plus an extended distribution of
gas present in the channels near M31's systemic velocity (-300 \kms).
Images for the following heliocentric velocities are shown: -656 (top
left), -622 ... -181 \kms (bottom right).\label{fig1}}
\end{figure*}

\clearpage 

\begin{figure}
\epsscale{0.8}
\plotone{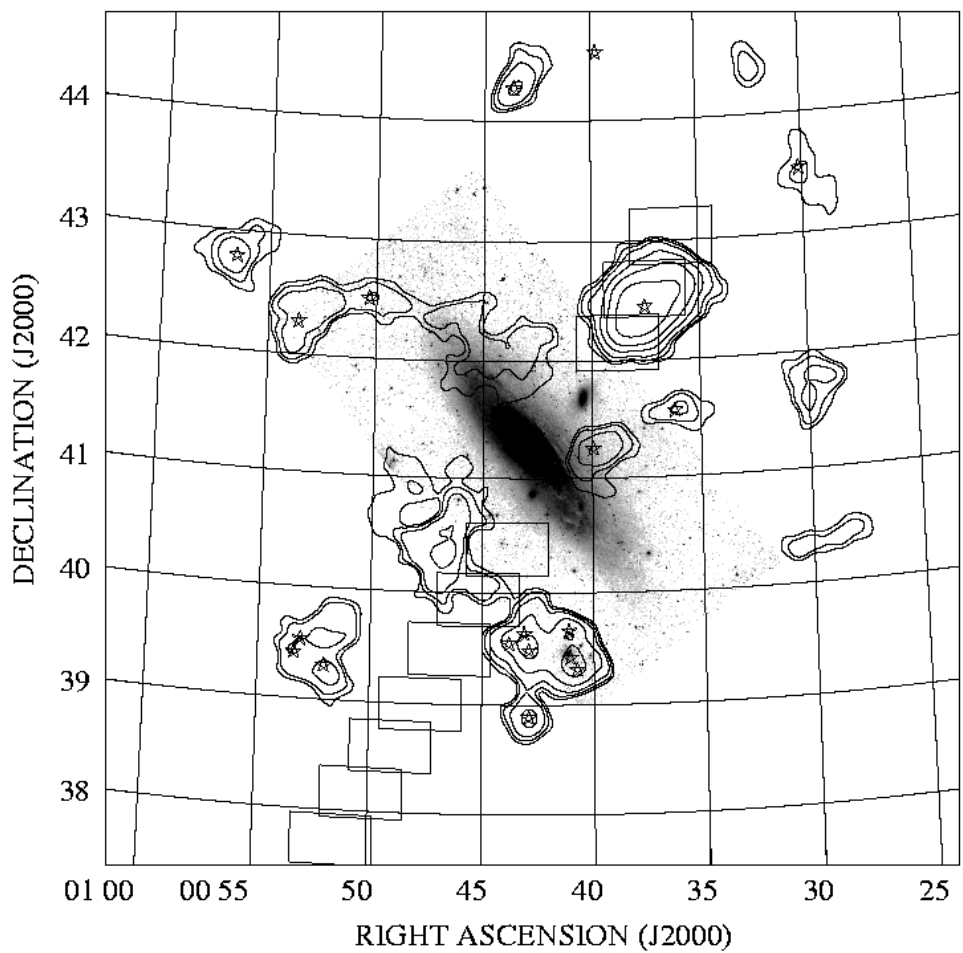}
\caption{Total column density for discrete and diffuse high-velocity
\ion{H}{1} in the M31 GBT field, after masking emission from
Andromeda's inclined, rotating disk.  Contours were evaluated at (3
kpc, 72 \kms) resolution and rendered at 0.5, 1, 2, 10, and
$20\times10^{18}$~cm$^{-2}$, then overlaid on a V band image of M31.
The position of each discrete cloud (or distinct substructure) is
marked using a star symbol.  We indicate the location of the
metal-rich stellar stream analyzed by Ferguson et al. (2002), by
plotting the imaging fields of McConnachie et al. (2003).  \ion{H}{1}
emission from NGC205 has been masked.\label{fig2}}
\end{figure}

\clearpage

\begin{figure}
\epsscale{0.8}
\plotone{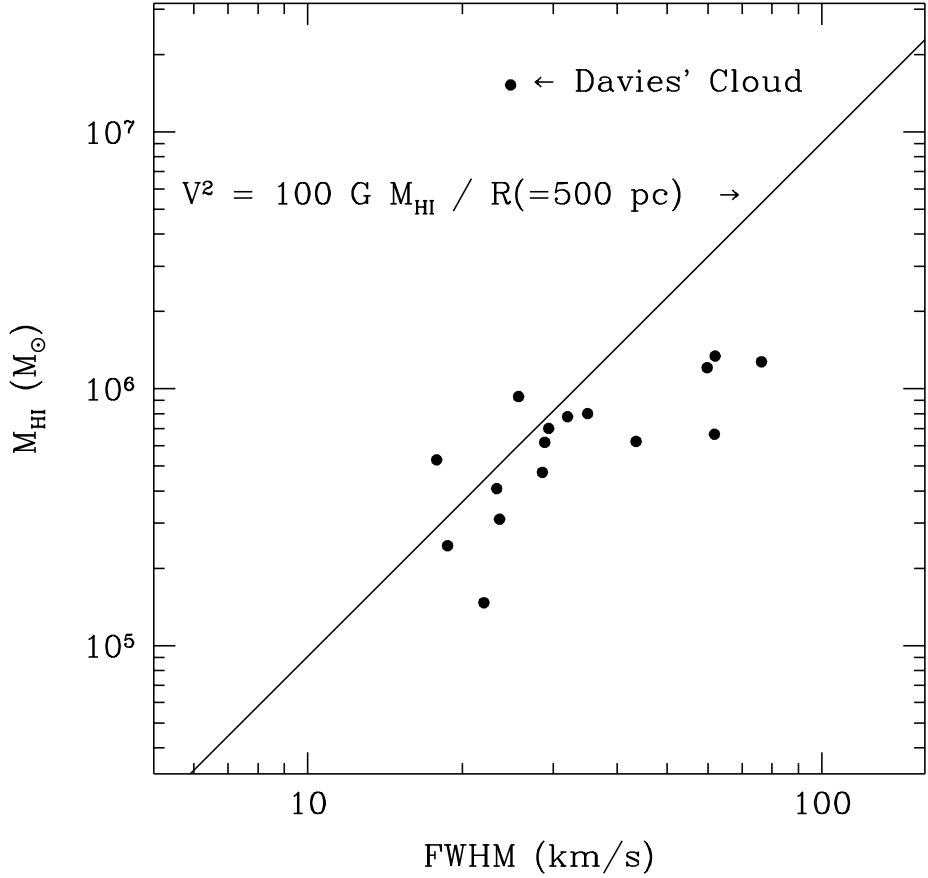}
\caption{Observed \ion{H}{1} mass versus FWHM line-width for discrete
 clouds near M31.  Line-width was measured using Gaussian fits to
 flux-weighted average spectra determined over the extent of each
 cloud.  We have also plotted the anticipated relationship between
 \ion{H}{1} mass and line-width for gravitationally confined clouds.
 To first order, our FWHM measurements appear consistent with the
 hypothesis that the objects are dark matter dominated, assuming a
 dark matter to \ion{H}{1} mass ratio of 100:1 and a characteristic
 size of 0.5 kpc for each \ion{H}{1} cloud core.  The later assumption
 is supported by our WSRT detection of resolved, high column-density
 cores (N$_{HI} \sim 10^{19}-10^{20}$~cm$^{-2}$) within many of the
 more centrally-located clouds.
\label{fig3}}
\end{figure}



\end{document}